# Real-space Hamiltonian method for low-dimensional semiconductor heterostructures

Yong-Hee Cho and Alexey Belyanin[*]

Department of Physics & Astronomy, Texas A&M University, College Station, Texas 77840-4242

## ABSTRACT

We present a new method for calculating electronic states in low-dimensional semiconductor heterostructures, which is based on the real-space Hamiltonian in the envelope function approximation. The numerical implementation of the method is extremely simple; all subband energy levels and envelope functions are directly obtained by a single evaluation of the heterostructure Hamiltonian matrix. We test the method in the 6- and 8-band **k** • **p** models as well as in a simple parabolic one-band model and demonstrate its great accuracy. The method can be straightforwardly generalized to a general n-band **k** • **p** model. We describe three different approaches within the method which make it possible to investigate the origin and removal of the spurious or unphysical solutions, which has long been an important issue in the community.

## I. INTRODUCTION

The development of semiconductor epitaxial growth technologies such as the molecular beam epitaxy (MBE)[1] and the metal-organic chemical vapor deposition (MOCVD)[2] has enabled the fabrication of high-quality nanostructures, in which layer thicknesses are controlled with one monolayer precision and electrons experience quantum confinement in one, two, or three dimensions.[3] The methods of calculating confined electron states, particularly in semiconductor layered heterostructures, are more complex than those in the bulk materials since the momentum component in the confinement direction (or the growth direction) becomes an operator, allowing only discrete eigen-energies in that direction instead of continuous ones. It requires solving coupled high-order partial differential equations depending on the number of bands included. This is usually handled by complicated numerical techniques as well as large computing resources. Therefore, it is important to find a simple, accurate, and efficient methodology.

A number of methods in the real space[4-7] as well as in the momentum space[8-10] have been developed for calculating eigenstates in coupled low dimensional semiconductor heterostructures. Many of them suffer from unphysical spurious solutions, and much effort have been spent on trying to remove these artifacts in various ways[9-18] even without being restricted[13, 18-20] to the k • p theory.[21-24] However, it seems that there are no universal methods for removing spurious solutions until now. They are usually applicable for particular or simplified band models, or for specific numerical models, and it is sometimes required to change the k • p Hamiltonian matrix or band parameters. The methods are not compatible to one another.

The spurious solutions are usually referred to as eigen solutions that are located in the middle of the band gap or that are fast oscillating envelope functions. It is heuristically known that multi-band Hamiltonians, expanded up to the second order in terms of the confined momentum component, generate such spurious solutions upon simultaneously seeking eigenstates in both the conduction and the valence bands. This happens regardless of the simplicity of the k • p band models. Often, the origin is attributed to large k values outside the first Brillouin zone edge.

In the momentum space, the cut-off method[9] has been recently suggested for removing fast oscillating envelope function-type spurious solutions in the plane wave expansion for confined states. It truncates the wave number vector at the cut-off value which is determined by the conduction band bending and is much smaller than the edge of the first Brillouin zone. The method is based on a bulk 8-band structure, in which the contribution due to the explicit inclusion of the conduction band had not been subtracted from the valence band parameters. In other words, the Luttinger parameters were not modified properly. Consequently, the conduction band is bowed down into the band gap as the wave vector increases, which is obviously unphysical. The basic idea of introducing a cutoff to remove spurious solutions has been employed in other papers.[13, 18]

The correct interface boundary condition, which connects wells and barriers at abrupt jump interfaces in heterostructures, is still being disputed.[25-31] Here, we follow the ideology of the Fourier grid Hamiltonian (FGH) method[32], in which explicit boundary conditions are not necessary. Therefore, the uncertainty in the choice of the interface boundary conditions does not affect the eigen solutions in semiconductor heterostructure problems within our method. Such approach to the heterostucture problem has been previously used in Ref.[8] in the momentum space instead of that in the real space as in our present work.

The FGH method[32] uses the forward and backward Fourier transformation, the variational method, and the fact that natural representations for the kinetic and the potential energies are in the momentum and the coordinate space, respectively. The resulting Hamiltonian for a bound system in a simple one-dimensional Schrödinger equation forms an N×N square matrix, in which N is equally discretized number of grid points in the coordinate space. The method has been further developed in various ways.[33-38]

In this paper, we develop three different approaches for calculating the eigenstates in a heterostructure. First, starting from the machinery of the FGH method, we set up the formalism for the three cases in the one-band model, and then they are extended to the n-band **k • p** model based on the envelope function approximation (EFA). The differences between the approaches come from different kinds of approximations in dealing with the quasiparticle momentum integrals, which unavoidably appear due to the coordinate-dependent band parameters. These approximations have a simple physical interpretation and their analysis helps us to pinpoint the origin of spurious solutions and to remove them.

## II. ONE BAND MODEL

### A. Two dimensional heterostructure Hamiltonian based on the Fourier grid Hamiltonian method

It is usually assumed that energy bands at a heterojunction between two different bulk semiconductors have a sharp discontinuity which can be determined from the empirically known offset between the positions of the valence band edges of the two bulk materials. InSb provides a zero reference of the valence band offset, from which those for the other materials are determined.[39] The resulting band edge profiles in a heterojunction look step-function-like and therefore acquire the coordinate dependence. This is true not only for the band edges but also for other band parameters such as the Luttinger parameters, the Kane parameter, and strain parameters because band parameters used in heterostructure problems are assumed to be the same as the parameters of bulk materials in the EFA.

In the parabolic one band model, the band edge effective masses for quasi-particles also show abrupt jumps and coordinate-dependence at heterojunctions. Therefore, in the Schrödinger equation for the one band model (Eq. (1)), one needs to decide how to write the kinetic energy term, which contains the inverse coordinate-dependent mass and the momentum which becomes a differential operator in the coordinate representation. The usual choice is to write the kinetic term in the symmetrized way that keeps the Hamiltonian Hermitian[40]:

$$\hat{H} = \frac{\hbar^2 \hat{k}_z^2}{2m^*(\hat{z})} + V(\hat{z}) = \hat{k} B(\hat{z}) \hat{k} + V(\hat{z}) \qquad (1)$$

where the confinement direction is assumed to be the z-direction, $B(\hat{z}) = \hbar^2/2m^*(\hat{z})$, $\hbar \hat{k}_z$ is the momentum operator, the notation for $\hat{k}_z$ is simplified to $\hat{k}$, and $V(\hat{z})$ is a quantum well potential in a single band,.

We first adopt the machinery of the Fourier grid Hamiltonian (FGH) method[32] to derive the Hamiltonian for semiconductor heterostructures. In (1), $B(\hat{z})$ and $V(\hat{z})$ are represented in the coordinate basis, and $\hat{k}_z$ in the momentum basis. These are most natural choices because of the immediate diagonalization of eigenvalues in each representation. Then, (1) can be expressed as

$$\langle z|\hat{H}|z'\rangle = \langle z|\hat{k}B(\hat{z})\hat{k} + V(\hat{z})|z'\rangle \qquad (2)$$

$$= \int_{-\infty}^{\infty}\int_{-\infty}^{\infty}\int_{-\infty}^{\infty}\int_{-\infty}^{\infty}\int_{-\infty}^{\infty}\int_{-\infty}^{\infty} \langle z|k\rangle\langle k|\hat{k}|k'\rangle\langle k'|z''\rangle\langle z''|B(\hat{z})|z'''\rangle\langle z'''|k''\rangle\langle k''|\hat{k}|k'''\rangle\langle k'''|z'\rangle dk dk' dk'' dk''' dz'' dz'''$$

$$+ \langle z|V(\hat{z})|z'\rangle \qquad (3)$$

where the completeness relations for coordinate and momentum bases were used, i.e.,

$$\hat{I}_z = \int_{-\infty}^{\infty} |z\rangle\langle z| dz, \quad \hat{I}_k = \int_{-\infty}^{\infty} |k\rangle\langle k| dk$$

Assuming the plane wave basis when projecting the momentum space on the coordinate space

$$\langle z|k\rangle = \frac{1}{\sqrt{2\pi}} e^{ikz}$$

and using the orthogonality of the basis along with eigenvalue equations such that

$$\hat{k}|k'\rangle = k'|k'\rangle, \quad B(\hat{z})|z'''\rangle = B(z''')|z'''\rangle$$

(3) can be simplified to (4) leaving only three integrals in the kinetic energy term:

$$\langle z|\hat{H}|z'\rangle = \frac{1}{(2\pi)^2} \int_{-\infty}^{\infty}\int_{-\infty}^{\infty}\int_{-\infty}^{\infty} k e^{ik(z-z'')} B(z'') e^{ik'(z''-z')} k' \, dk \, dk' \, dz'' + V(z)\delta(z-z') \quad (4)$$

Note that as a result of the coordinate dependent band parameter, i.e., B(z), integrations in terms of two different quasi-particle momenta appear as the symmetric form in (4). Since the bases are spanning over the coordinates of a whole quantum well system, the explicit treatment of boundary conditions are not required. Instead they are implicitly included and automatically fulfilled. This is also true when the band parameter is not dependent of coordinates. Starting from (4), three different approaches depending on approximations of integrals in (4) will be shown through this section and section II. B ~ II. C.

For the numerical treatment in the FGH method, the integration in Eq. (4) is straightforwardly done by the discretization with an equal grid length, $\Delta z = L/(N-1)$, which has the Fourier reciprocal relation with momentum as $\Delta k = 2\pi/(N\Delta z)$. Here, L and N are the total system length in the coordinate space and the total odd number of grid points, respectively. Then, we change integral variables and coordinate bases to discrete forms in (4) such that $k = \alpha \Delta k$, $k' = \beta \Delta k$, $z = (p-1)\Delta z$, $z' = (q-1)\Delta z$, and $z'' = (s-1)\Delta z$, where $-m \leq \alpha, \beta \leq m$ (=(N-1)/2), and p, q, and s are 1, 2, 3,…, N. Integrals are now replaced by summations. Then, the kinetic energy term in (4) becomes the following expression:

$$\frac{1}{(2\pi)^2} \sum_{s=1}^{N} \left[ B((s-1)\Delta z)\Delta z \sum_{\beta=-m}^{m} (\beta \Delta k)(\Delta k) e^{i(\beta \Delta k)(p-s)\Delta z} \sum_{\alpha=-m}^{m} e^{i(\alpha \Delta k)(s-q)\Delta z} (\alpha \Delta k)(\Delta k) \right]$$

$$= \frac{1}{\Delta z} \frac{4(\Delta k)^2}{N^2} \sum_{s=1}^{N} \left[ B((s-1)\Delta z) \left( \sum_{\beta=1}^{m} \beta \cos\left[\frac{2\pi\beta(p-s)}{N}\right] \right) \left( \sum_{\alpha=1}^{m} \alpha \cos\left[\frac{2\pi\alpha(s-q)}{N}\right] \right) \right] \quad (5)$$

where we have used that $\alpha, \beta = 0$ do not contribute to the summations. The potential energy term in (4) can be discretized in the same way as the kinetic energy term, and resulting equation can be written as

$$V((p-1)\Delta z)\delta((p-q)\Delta z)$$
$$\approx \frac{1}{\Delta z} V((p-1)\Delta z)\delta_{pq} \quad (6)$$

where the following property of the delta function[41] was used:

$$\delta(\Delta z(p-q)) = \frac{1}{\Delta z}\delta(p-q) \approx \frac{1}{\Delta z}\delta_{pq}, \quad (\Delta z > 0) \quad (7)$$

By putting (5) and (6) together and applying the variational method,[32] which cancels the $1/\Delta z$ factor in (5) and (6), we finally obtain the Hamiltonian matrix elements for quantum well heterostructures in the one band model based on the Fourier grid Hamiltonian method:

$$H(p,q) = \frac{4(\Delta k)^2}{N^2} \sum_{s=1}^{N} \left[ B((s-1)\Delta z) \left( \sum_{\beta=1}^{m} \beta \cos\left[\frac{2\pi\beta(p-s)}{N}\right] \right) \left( \sum_{\alpha=1}^{m} \alpha \cos\left[\frac{2\pi\alpha(s-q)}{N}\right] \right) \right] + V((p-1)\Delta z)\delta_{pq}$$

(8)

Note that the Hamiltonian matrix is real and symmetric upon exchanging basis indices p and q, ensuring that it is Hermitian. The two summations in terms of α and β are independent of each other. However, the summation over s is connected with the other summations.

Using a standard eigenvalue equation solver, the eigenvalues ($E_n$) (or subband energy levels) and the eigenfunctions ($f_n$) (or envelope functions) are readily obtained by solving $Hf_n = E_n f_n$. Each eigenfunction is normalized by fulfilling the following condition:

$$\Delta z \sum_{m=1}^{N} f_n((m-1)\Delta z) = 1$$

### B. Modified formalism with improved accuracy

In the previous section, the final form of a heterostructure Hamiltonian (8) in the one band model has been obtained by directly discretizing Eq. (4), following the original strategy of the Fourier grid Hamiltonian method. All three integrals in (4) have been approximated by the discretization, which inevitably generates errors in determining the subband energy levels. In this section we show how the method can be improved by the analytical evaluation of the quasi-particle momentum integrals. It is possible since the infinite integral range is practically cut off to be finite, determined by the Fourier reciprocal relation as was done in (5).

Therefore, the integral over k in Eq. (4) can be rewritten as

$$\int_{-k_m}^{k_m} k e^{ik(z-z'')} dk = -2ik_m \frac{\partial}{\partial z}\left[\text{sinc}(k_m(z-z''))\right] \tag{9}$$

$$= \begin{cases} -2i\left[\dfrac{k_m \cos(k_m(z-z''))}{(z-z'')} - \dfrac{\sin(k_m(z-z''))}{(z-z'')^2}\right] & \text{when } z \neq z'' \\ 0 & \text{when } z = z'' \end{cases} \tag{10}$$

Here sinc is an unnormalized sinc function, $k_m = m\Delta k$, where m and $\Delta k$ are defined as in (5). The integral over k' can be calculated similarly. Consequently, the kinetic energy part of (4) becomes

$$\frac{4}{(2\pi)^2}\int_0^L B(z'')\left[\frac{k_m \cos(k_m(z-z''))}{(z-z'')} - \frac{\sin(k_m(z-z''))}{(z-z'')^2}\right]\left[\frac{k_m \cos(k_m(z'-z''))}{(z'-z'')} - \frac{\sin(k_m(z'-z''))}{(z'-z'')^2}\right]dz'' \quad (11)$$

where $z \neq z''$ and $z' \neq z''$. Now only a single integral exists as compared to (4). By the discretization as in (5) and the variational method used in (8), the modified heterostructure Hamiltonian of (8) with improved accuracy is given by

$$H(p,q) = \frac{4}{(2\pi)^2}(\Delta z)^2 \sum_{s=1}^{N}\left[B((s-1)\Delta z)\left(\frac{k_m \cos\left(\frac{2\pi}{N}m(p-s)\right)}{(p-s)\Delta z} - \frac{\sin\left(\frac{2\pi}{N}m(p-s)\right)}{(p-s)^2(\Delta z)^2}\right)\right.$$

$$\left.\times\left(\frac{k_m \cos\left(\frac{2\pi}{N}m(q-s)\right)}{(q-s)\Delta z} - \frac{\sin\left(\frac{2\pi}{N}m(q-s)\right)}{(q-s)^2(\Delta z)^2}\right)\right] + V((p-1)\Delta z)\delta_{pq} \quad (12)$$

where the notations are same as above, and the Hamiltonian matrix is real and Hermitian again. Note that when p = s or q = s, the kinetic energy part in (12) is zero.

### C. Heterostructure Hamiltonian based on the delta function approach

In this section we rewrite Eq. (4) in terms of the delta functions in the real space and then discretize them. Starting from (4), in which no approximation has been used, the exponential terms are expressed as

$$ke^{ik(z-z'')} = \begin{cases} -i\frac{\partial}{\partial z}e^{ik(z-z'')} & (13) \\ i\frac{\partial}{\partial z''}e^{ik(z-z'')} & (14) \end{cases} \qquad k'e^{ik'(z''-z')} = \begin{cases} i\frac{\partial}{\partial z'}e^{ik'(z''-z')} & (15) \\ -i\frac{\partial}{\partial z''}e^{ik'(z''-z')} & (16) \end{cases}$$

Here, one might think that there could be several choices to replace the exponential terms since there are two possible derivatives for each. However, it is not true for the following reasons. First of all, we need to make sure that the Hamiltonian is still symmetric upon exchanging z and z' in the final form. Second, we would like to eliminate the integral with respect to z" in (4), eventually removing all integrals. To achieve this, each of the exponential terms should contain a z" derivative only once. The only possible choice which fulfills the above two conditions is to combine [(13) and (16)] and [(14) and (15)]. This is uniquely determined and excludes any other combinations. This becomes more transparent through the following derivation.

Using the above combinations, Eq. (4) can be now rewritten as

$$\langle z|\hat{H}|z'\rangle = \frac{-1}{2(2\pi)^2} \int_{-\infty}^{\infty} \left[ \left( \frac{\partial}{\partial z} \int_{-\infty}^{\infty} e^{ik(z-z'')} dk \right) B(z'') \left( \frac{\partial}{\partial z''} \int_{-\infty}^{\infty} e^{ik'(z''-z')} dk' \right) \right.$$
$$\left. + \left( \frac{\partial}{\partial z''} \int_{-\infty}^{\infty} e^{ik(z-z'')} dk \right) B(z'') \left( \frac{\partial}{\partial z'} \int_{-\infty}^{\infty} e^{ik'(z''-z')} dk' \right) \right] dz'' + V(z)\delta(z-z') \quad (17)$$

Note that the derivatives do not act on B (z"). The integrals in terms of quasi-particle momenta k and k' can be exactly evaluated using the integral form of the delta function,[41] i.e.,

$$\delta(z-z') = \frac{1}{2\pi} \int_{-\infty}^{\infty} e^{ik(z-z')} dk$$

Then, the kinetic energy term in (17) becomes

$$\frac{-1}{2} \int_{-\infty}^{\infty} \left[ \left( \frac{\partial}{\partial z} \delta(z-z'') \right) \left( \frac{\partial}{\partial z''} \delta(z''-z') \right) B(z'') + \left( \frac{\partial}{\partial z''} \delta(z-z'') \right) \left( \frac{\partial}{\partial z'} \delta(z''-z') \right) B(z'') \right] dz'' \quad (18)$$

Using the relation for the derivative of the delta function,

$$-\frac{\partial f(z')}{\partial z'} = \int_{-\infty}^{\infty} \frac{\partial \delta(z-z')}{\partial z} f(z) dz$$

formula (18) can be further simplified by explicitly evaluating the integral with respect to z", i.e.,

$$\frac{1}{2}\left[\frac{\partial}{\partial z'}\left[\left(\frac{\partial}{\partial z}\delta(z-z')\right)B(z')\right]+\frac{\partial}{\partial z}\left[\left(\frac{\partial}{\partial z'}\delta(z-z')\right)B(z)\right]\right] \qquad (19)$$

By putting the kinetic energy together with the potential energy term, (17) can be now read as

$$\langle z|H|z'\rangle = \frac{1}{2}\left[\frac{\partial}{\partial z'}\left[\left(\frac{\partial}{\partial z}\delta(z-z')\right)B(z')\right]+\frac{\partial}{\partial z}\left[\left(\frac{\partial}{\partial z'}\delta(z-z')\right)B(z)\right]\right]+V(z)\delta(z-z') \qquad (20)$$

This is the core equation of the delta function approach for layered heterostructure problems in the one band model. Eq. (20) can be compared to Eq. (4). Now all integrals have disappeared, and instead the Hamiltonian consists of derivatives of the delta functions and solely depends on the band parameters in the real space. It should be noted that Eq. (20) has been obtained without any approximation starting from (1) and the integrals for k and k' have been evaluated out in the infinite range rather than within the effective cut-off values as in (8) and (12). This difference plays an important role in investigating the spurious solutions in later sections. Note that Eq. (20) is real and symmetric with respect to the coordinates z and z'. This is the result of the symmetrization which was discussed above and which was required due to the basis-dependent band parameter B (z) as well as non-zero off-diagonal elements in the kinetic energy matrix. Such additional symmetrization was not necessary in (12) since the argument of the band parameter B(z") is for summation but not for Hamiltonian basis.

When B (z) is constant, B (z) =B, Eq. (20) is reduced to

$$\langle z|\hat{H}|z'\rangle = B\frac{\partial}{\partial z'}\left[\left(\frac{\partial}{\partial z}\delta(z-z')\right)\right]+V(z)\delta(z-z') \qquad (21)$$

The readers can intuitively consider (21) as the original Schrödinger equation in which the envelope function has been replaced by the delta function with a minus sign in the kinetic energy term, associating two coordinate bases, not just one.

Before discretizing Eq. (20) for the numerical treatment, we replace the derivative of the delta function by the alternative definition based on the finite difference form:

$$\frac{\partial}{\partial z}\delta(z-z') = \lim_{h\to 0}\frac{\delta(z+h-z')-\delta(z-z')}{h} \qquad (22)$$

Then, the kinetic energy term in (20) becomes

$$\frac{1}{2}\lim_{h\to 0}\left(\frac{B(z'+h)\left(\frac{\delta(z+h-z'-h)-\delta(z-z'-h)}{h}\right)-B(z')\left(\frac{\delta(z+h-z')-\delta(z-z')}{h}\right)}{h}\right.$$

$$\left.+\frac{B(z+h)\left(\frac{\delta(z+h-z'-h)-\delta(z+h-z')}{h}\right)-B(z)\left(\frac{\delta(z-z'-h)-\delta(z-z')}{h}\right)}{h}\right)$$

$$= \frac{1}{2}\lim_{h\to 0}\left[\frac{1}{h^2}\left(\delta(z-z')[B(z'+h)+B(z')+B(z+h)+B(z)]\right.\right. \qquad (23)$$

$$\left.\left.-\delta(z-z'-h)[B(z'+h)+B(z)]-\delta(z-z'+h)[B(z+h)+B(z')]\right]\right]$$

Note that (23) is still an exact expression without any approximation as long as h infinitesimally goes to zero.

Now we discretize and approximate (23) by the equal grid length in the coordinate space, similarly to Eq. (5). By substituting $z=(p-1)\Delta z$, $z'=(q-1)\Delta z$, $h=\Delta z=L/(N-1)$, where p and q are 1, 2, 3,..., N, and N is the total number of coordinate grid points, formula (23) becomes

$$\frac{1}{2(\Delta z)^2}(\delta((p-q)\Delta z)[B(q\Delta z)+B((q-1)\Delta z)+B(p\Delta z)+B((p-1)\Delta z)]$$

$$-\delta((p-q-1)\Delta z)[B(q\Delta z)+B((p-1)\Delta z)]-\delta((p-q+1)\Delta z)[B(p\Delta z)+B((q-1)\Delta z)])$$

Using the delta function property of Eq. (7), we obtain the discretized form of the kinetic energy term:

$$\frac{1}{2(\Delta z)^2}\left(\frac{1}{\Delta z}\delta_{pq}[B(q\Delta z)+B((q-1)\Delta z)+B(p\Delta z)+B((p-1)\Delta z)]\right.$$
$$\left.-\frac{1}{\Delta z}\delta_{p,q+1}[B(q\Delta z)+B((p-1)\Delta z)]-\frac{1}{\Delta z}\delta_{p,q-1}[B(p\Delta z)+B((q-1)\Delta z)]\right)$$
$$=\frac{1}{2(\Delta z)^2}\frac{1}{\Delta z}\left(\delta_{pq}[B(q\Delta z)+B((q-1)\Delta z)+B(p\Delta z)+B((p-1)\Delta z)]\right.$$
$$\left.-\delta_{p,q+1}[B(q\Delta z)+B((p-1)\Delta z)]-\delta_{p,q-1}[B(p\Delta z)+B((q-1)\Delta z)]\right) \quad (24)$$

The potential term in (20) is also discretized to the same form as in (7). Putting it together with (24) and using the variational method as in (5), (6) and (8), we obtain the final form of the heterostructure Hamiltonian in the delta function approach:

$$H(p,q)=\frac{1}{2(\Delta z)^2}\left(\delta_{pq}[B(q\Delta z)+B((q-1)\Delta z)+B(p\Delta z)+B((p-1)\Delta z)]\right.$$
$$\left.-\delta_{p,q+1}[B(q\Delta z)+B((p-1)\Delta z)]-\delta_{p,q-1}[B(p\Delta z)+B((q-1)\Delta z)]\right)+V((p-1)\Delta z)\delta_{pq} \quad (25)$$

The eigen solutions of (25) can be obtained in the same way as described in section II. A. Compared to the previous formalism in Eqs. (8) and (12), the kinetic energy part has been dramatically simplified. It is only necessary to know tridiagonal terms, originated from the finite difference form of the derivative of the delta function without any summations. The total number of grid points N does not have to be odd, unlike in Eq. (8).

When B (z) is a constant in (25), i.e., B (z) = B, the symmetric form for B (z) is no longer necessary, leading to a much simpler form of the Hamiltonian:

$$H(p,q)=\frac{B}{(\Delta z)^2}\left[2\delta_{pq}-\delta_{p,q+1}-\delta_{p,q-1}\right]+V((p-1)\Delta z)\delta_{pq} \quad (26)$$

Note that Eq. (26) can be used for any quantum systems with one-dimensional confinement potential and a constant mass. The compactness and simplicity of the Hamiltonian (25) and (26) differentiate it from the other methods. The superior calculation speed is obvious due to the sparse Hamiltonian matrix elements, which can be immediately defined without any calculus.

The delta function method should not be confused with the finite difference method (FDM)[7] which also employs the tridiagonal matrix. Nevertheless, they are different methods for the following obvious reasons: In the FDM, (i) boundary conditions should be explicitly taken into account upon constructing system equations. (ii) the tridiagonal matrix is not a Hamiltonian matrix. In fact, it comes from the recurrence relation of envelope functions directly discretized at the very beginning from the Schrödinger equation along with boundary conditions at heterojunctions; (iii) the tridiagonal matrix includes the eigenvalues, which have to be solved for; (iv) the matrix inversion is required to obtain envelope functions.

In the following section, we extend the formalism of our method to general multiband **k • p** models based on the EFA.

### III. REAL-SPACE HETEROSTRUCTURE HAMILTONIAN IN n-BAND k • p MODELS

Compared to a parabolic energy dispersion as in (1), the actual band structure of semiconductors is much more complicated since each band is coupled with other bands, leading to the non-parabolicity and the anisotropy. Therefore, except very near the high symmetry points at the band extrema, the parabolic approximation with a constant band-edge effective mass is generally not adequate. There is an approximate recipe to include the non-parabolic correction[42, 43] within a one band model through the energy-dependent effective mass or an additional term to the kinetic energy which is of the fourth order in k$_z$. However, its use should be still limited by the close vicinity of the band extrema by definition.

Including band-to-band interactions for an accurate band structure by explicitly solving the coupled high order partial differential equations is essential especially when dealing with low-dimensional confinement, electron states with high in-plane momenta or large free carrier densities. For the transition from bulk to heterostructure problems, each basis of the N coordinate space is acted on a general n-band bulk Hamiltonian $H_{n \times n}$. The resulting matrix is a nN×nN heterostructure Hamiltonian shown in (27).

$$\begin{pmatrix} \langle 1|H_{n\times n}|1\rangle & \langle 1|H_{n\times n}|2\rangle & \cdots & \langle 1|H_{n\times n}|N\rangle \\ \langle 2|H_{n\times n}|1\rangle & \langle 2|H_{n\times n}|2\rangle & \cdots & \langle 2|H_{n\times n}|N\rangle \\ \vdots & \vdots & \ddots & \vdots \\ \langle N|H_{n\times n}|1\rangle & \langle N|H_{n\times n}|2\rangle & \cdots & \langle N|H_{n\times n}|N\rangle \end{pmatrix} \begin{pmatrix} \overline{F}_{1n} \\ \overline{F}_{2n} \\ \vdots \\ \overline{F}_{Nn} \end{pmatrix} = E \begin{pmatrix} \overline{F}_{1n} \\ \overline{F}_{2n} \\ \vdots \\ \overline{F}_{Nn} \end{pmatrix} \qquad (27)$$

where

$$\overline{F}_{in} = \begin{pmatrix} f_i^1 \\ f_i^2 \\ \vdots \\ f_i^n \end{pmatrix}$$

and E is a real constant eigenvalue. For the whole system, eigenvalues and eigenfunctions are also nN×nN matrices. $\overline{F}_{in}$ is a column vector of length n for i$^{th}$ coordinate basis.

However, Eq. (27) is not a convenient form to directly apply the formalism obtained in the previous sections. Therefore, using the linear algebra, we rewrite (27) as

$$\begin{pmatrix} \langle i|H_{11}|j\rangle & \langle i|H_{12}|j\rangle & \cdots & \langle i|H_{1n}|j\rangle \\ \langle i|H_{21}|j\rangle & \langle i|H_{22}|j\rangle & \cdots & \langle i|H_{2n}|j\rangle \\ \vdots & \vdots & \ddots & \vdots \\ \langle i|H_{n1}|j\rangle & \langle i|H_{n2}|j\rangle & \cdots & \langle i|H_{nn}|j\rangle \end{pmatrix} \begin{pmatrix} F_{1N} \\ F_{2N} \\ \vdots \\ F_{nN} \end{pmatrix} = E \begin{pmatrix} F_{1N} \\ F_{2N} \\ \vdots \\ F_{nN} \end{pmatrix} \quad (28)$$

where

$$F_{nN} = \begin{pmatrix} f_1^n \\ f_2^n \\ \vdots \\ f_N^n \end{pmatrix}.$$

Here i and j run over 1 to N, $\langle i|H_{\alpha\beta}|j\rangle$ is an N×N square matrix, n is a bulk band index, and $F_{nN}$ is a column vector of length N, representing the envelope function for $n^{th}$ band. Eq. (28) shows that a heterostructure Hamiltonian is constructed such that each bulk Hamiltonian matrix element $H_{\alpha\beta}$ is transformed to an N×N square block matrix, $\langle i|H_{\alpha\beta}|j\rangle$.

In this paper, we restrict the highest order of momentum to the second order in bulk Hamiltonian matrix elements. A usual 8-band k • p model (see Appendix) conforms to this category. Therefore, the bulk Hamiltonian matrix elements contain zeroth order, linear, and quadratic terms with respect to the confined wave number $k_z$.

In the one band case, we have already shown how the second order term with respect to $k_z$ is transformed to the heterostructure case with N grid points in the coordinate space for three different approaches. Therefore, each of them can be readily applied to the conversion of bulk multi-band Hamiltonian matrix elements to the heterostructure case. They are explicitly rewritten again from sections II. A ~ II. C as

$$H_{\alpha\beta}(k_z^2) \Rightarrow \frac{4(\Delta k)^2}{N^2} \sum_{s=1}^{N} \left[ B_{\alpha\beta}((s-1)\Delta z) \left( \sum_{a=1}^{m} a \cos\left[\frac{2\pi a(p-s)}{N}\right] \right) \left( \sum_{b=1}^{m} b \cos\left[\frac{2\pi b(s-q)}{N}\right] \right) \right] \quad (29)$$

$$H_{\alpha\beta}(k_z^2) \Rightarrow \frac{4}{(2\pi)^2} (\Delta z)^2 \sum_{s=1}^{N} \left[ B_{\alpha\beta}((s-1)\Delta z) \left( \frac{k_m \cos\left(\frac{2\pi}{N} m(p-s)\right)}{(p-s)\Delta z} - \frac{\sin\left(\frac{2\pi}{N} m(p-s)\right)}{(p-s)^2(\Delta z)^2} \right) \right.$$

$$\left. \times \left( \frac{k_m \cos\left(\frac{2\pi}{N} m(q-s)\right)}{(q-s)\Delta z} - \frac{\sin\left(\frac{2\pi}{N} m(q-s)\right)}{(q-s)^2(\Delta z)^2} \right) \right] \quad (30)$$

$$H_{\alpha\beta}(k_z^2) \Rightarrow \frac{1}{2(\Delta z)^2} \left( \delta_{pq} \left[ B_{\alpha\beta}(q\Delta z) + B_{\alpha\beta}((q-1)\Delta z) + B_{\alpha\beta}(p\Delta z) + B_{\alpha\beta}((p-1)\Delta z) \right] \right.$$
$$\left. - \delta_{p,q+1} \left[ B_{\alpha\beta}(q\Delta z) + B_{\alpha\beta}((p-1)\Delta z) \right] - \delta_{p,q-1} \left[ B_{\alpha\beta}(p\Delta z) + B_{\alpha\beta}((q-1)\Delta z) \right] \right) \quad (31)$$

where $H_{\alpha\beta}$ is a bulk **k · p** Hamiltonian matrix element, $B_{\alpha\beta}$ is the corresponding bulk band parameter, and the rest of the notations is the same as in the previous sections.

It remains to derive the linear terms, $B_{\alpha\beta}(\hat{z})\hat{k}_z$ and zeroth order terms to write a complete heterostructure Hamiltonian in the multiband case. Basically, the procedure of the derivation for the linear term is same as shown above for the quadratic term. In the coordinate space, $B_{\alpha\beta}(\hat{z})\hat{k}_z$ can be expressed as

$$\frac{1}{2} \langle z | B_{\alpha\beta}(\hat{z})\hat{k} + \hat{k} B_{\alpha\beta}(\hat{z}) | z' \rangle \quad (32)$$

where $\hat{k}_z$ is simplified to $\hat{k}$, and $B_{\alpha\beta}(\hat{z})\hat{k}_z$ has been symmetrized to be Hermitian. Following the procedure of going from (2) to (4), formula (32) becomes

$$\frac{1}{4\pi} \left[ (B_{\alpha\beta}(z) + B_{\alpha\beta}(z')) \int_{-\infty}^{\infty} e^{ik(z-z')} k \, dk \right] \quad (33)$$

In the quadratic term, two exponential terms were associated with two quasi-particle momenta, k and k'. In the linear term, only a single exponential term appears. Following sections II. A ~ II. C, the discretized forms of linear terms for the three approaches become

$$H_{\alpha\beta}(k_z) \Rightarrow \left(\frac{1}{N}\right)\Delta k \left[B_{\alpha\beta}((p-1)\Delta x) + B_{\alpha\beta}((q-1)\Delta x)\right] \sum_{a=1}^{m} a \cos\left[\frac{2a\pi(p-q)}{N}\right] \quad (34)$$

$$H_{\alpha\beta}(k_z) \Rightarrow -\frac{i}{2\pi}(\Delta z)\left[B_{\alpha\beta}((p-1)\Delta z) + B_{\alpha\beta}((q-1)\Delta z)\right]\left[\frac{k_m \cos\left(\frac{2\pi}{N}m(p-q)\right)}{(p-q)\Delta z} - \frac{\sin\left(\frac{2\pi}{N}m(p-q)\right)}{(p-q)^2(\Delta z)^2}\right],$$

$$(p \neq q) \quad (35)$$

$$H_{\alpha\beta}(k_z) \Rightarrow \frac{i}{4(\Delta z)}\left[B_{\alpha\beta}((p-1)\Delta z) + B_{\alpha\beta}((q-1)\Delta z)\right]\left(\delta_{p,q+1} - \delta_{p,q-1}\right) \quad (36)$$

where notations are the same as before. Contrary to the quadratic case, diagonal elements for linear terms are zeros in the block square matrix in Eqs. (35) and (36). In the delta function approach given by formula (36) one additional symmetrization has been performed by the linear combination of the two possible derivatives for $ke^{ik(x-x')}$ as shown in (9) and (10) to make the Hamiltonian Hermitian.

All terms that do not depend on $k_z$ or zeroth order terms with respect to $k_z$ in a bulk Hamiltonian matrix are transformed to an N×N diagonal matrix in the coordinate space regardless of the three different approaches in section II. A ~ II. C, i.e,

$$H_{\alpha\beta}((k_z)^0) \Rightarrow B_{\alpha\beta}\delta_{pq} \quad (37)$$

For example, in-plane momentum $\mathbf{k}_\parallel$-dependent terms and the potential energy terms correspond to this classification.

As we have shown above, the extension from the one band heterostructure problem to the multiband case is straightforward within our method once bulk k • p models are known. This extreme simplicity is the unique feature of the heterostructure Hamiltonian method.

In the next section, we show the numerical results of eigen solutions obtained by the heterostructure Hamiltonian method in the one band, the 6-valence band, and the 8-band k • p models. Particularly, in the case of the one band model, subband levels are compared to analytical results in single quantum well heterostructures.

## IV. NUMERICAL RESULTS AND SPURIOUS SOLUTIONS

### A. One band model

The numerical results in the one band model are shown in TABLE I and FIG. 1~3 for single quantum wells of $Ga_{0.47}In_{0.53}As$ surrounded by $Al_{0.48}In_{0.52}As$ barriers with 1 Å grid length and various quantum well widths. They are obtained by the three different methods derived in section II. A~II. C and compared to the analytical solutions.[24] The formalism that produces the most accurate result is the modified Fourier grid Hamiltonian method (MFGHM) based on the approach shown in the section II. B. The results obtained by the delta function method (DFM) in section II. C show nearly the same accuracy as those by the MFGHM. With the same grid length, both the MFGHM and the DFM are superior in the accuracy of eigen solutions to that of the shooting method, (for example, see Ref.[40]) which is widely used but limited to the one band model. The MFGHM provides an extreme accuracy along with simplicity. The formalism based on the original Fourier grid Hamiltonian method (FGHM) in section II. A shows the worst accuracy among the three approaches.

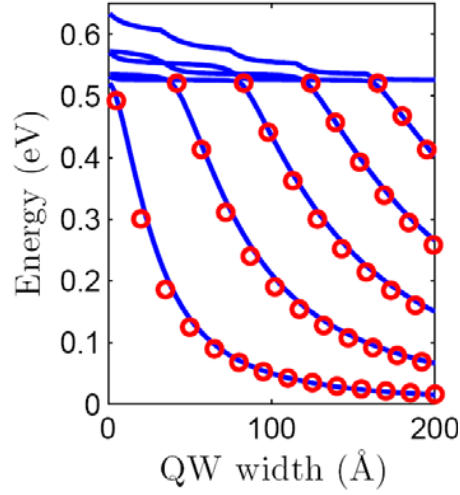

FIG. 1. (Color online) The confined subband energies obtained by the Fourier grid Hamiltonian method (blue solid lines) based on the section II. A are compared with analytical results (red circle) in $Ga_{0.47}In_{0.53}As$ / $Al_{0.48}In_{0.52}As$ single quantum wells as a function of the well width. Numerical values are shown in TABLE I. The discrepancy between the two cases is noticeable.

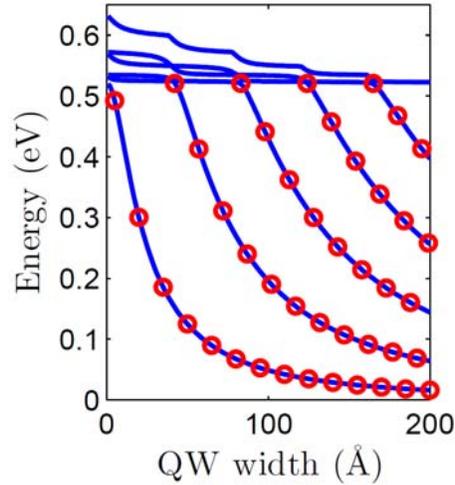

FIG. 2. (Color online) The confined subband energies obtained by the modified Fourier grid Hamiltonian method (blue solid line) based on the section II. B are compared with the analytic results (red circle) in $Ga_{0.47}In_{0.53}As$ / $Al_{0.48}In_{0.52}As$ single quantum wells as a function of the well

width. Numerical values are shown in TABLE I. The numerical and analytic solutions coincide for all eigenstates and well widths.

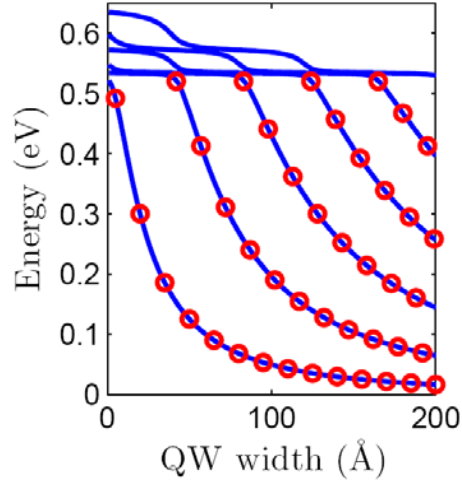

FIG. 3. (Color online) The confined subband energies obtained by the delta function method (blue solid lines) based on the section II. C are compared with the analytic results (red circle) in $Ga_{0.47}In_{0.53}As$ / $Al_{0.48}In_{0.52}As$ single quantum wells as a function of well width. Numerical values are provided in TABLE I.

TABLE I. Energies of confined eigenstates in a single $Ga_{0.47}In_{0.53}As$ quantum well surrounded by a $Al_{0.48}In_{0.52}As$ barrier. The numerical results obtained by the Fourier grid Hamiltonian method (FGHM), the modified Fourier grid Hamiltonian method (MFGHM), and the delta function method (DFM) in the one band model for grid length of 1 Å are compared with the analytic solutions for various well widths. The band parameters from Ref. 39 are used. The MFGHM shows the most accurate subband energies.

|  |  | Well width (Å) | | | | |
|---|---|---|---|---|---|---|
|  |  | 40 | 80 | 120 | 160 | 200 |
| $E_1$ (meV) | Analytical | 161.260 | 67.555 | 36.935 | 23.254 | 15.977 |
|  | FGHM | 182.438 | 70.903 | 36.627 | 22.196 | 14.840 |
|  | MFGHM | 161.200 | 67.537 | 36.927 | 23.250 | 15.975 |
|  | DFM | 161.148 | 67.513 | 36.916 | 23.244 | 15.971 |
| $E_2$ (meV) | Analytical | - | 269.970 | 148.172 | 93.212 | 63.998 |
|  | FGHM | - | 288.671 | 158.372 | 97.914 | 66.157 |
|  | MFGHM | - | 269.914 | 148.144 | 93.197 | 63.989 |
|  | DFM | - | 269.854 | 148.110 | 93.177 | 63.976 |
| $E_3$ (meV) | Analytical | - | - | 331.182 | 209.819 | 144.164 |
|  | FGHM | - | - | 343.663 | 219.863 | 150.220 |
|  | MFGHM | - | - | 331.140 | 209.791 | 144.147 |
|  | DFM | - | - | 331.056 | 209.744 | 144.118 |

The accuracy of confined subband levels can be further improved by decreasing grid lengths, eventually approaching the analytic solution as shown in TABLE II, calculated for a 20 Å single $Ga_{0.47}In_{0.53}As/Al_{0.48}In_{0.52}As$ quantum well .

TABLE II. Energy of the ground subband in a 20Å $Ga_{0.47}In_{0.53}As/Al_{0.48}In_{0.52}As$ single quantum well, calculated with the MFGM and the delta function method, and compared with the analytical solution. As grid length decreases, the numerical solutions approach the analytical solution.

| Grid length (Å) | Analytical solution (meV) | MFGH method (meV) | Delta function method (meV) |
|---|---|---|---|
| 1 |  | 300.1922 | 300.1366 |
| 0.5 | 300.3039 | 300.2757 | 300.2621 |
| 0.2 |  | 300.2994 | 300.2972 |
| 0.1 |  | 300.3028 | 300.3023 |

However, as shown in FIG. 4 (middle), the MFGHM generates fast oscillating continuum states i.e., the spurious solutions, which are the only subbands that are different from those in the DFM. Note that all confined electron states in FIG. 4 are free from spurious solutions and are not affected by the presence of the latter, as indicated in TABLE I and FIG. 2.

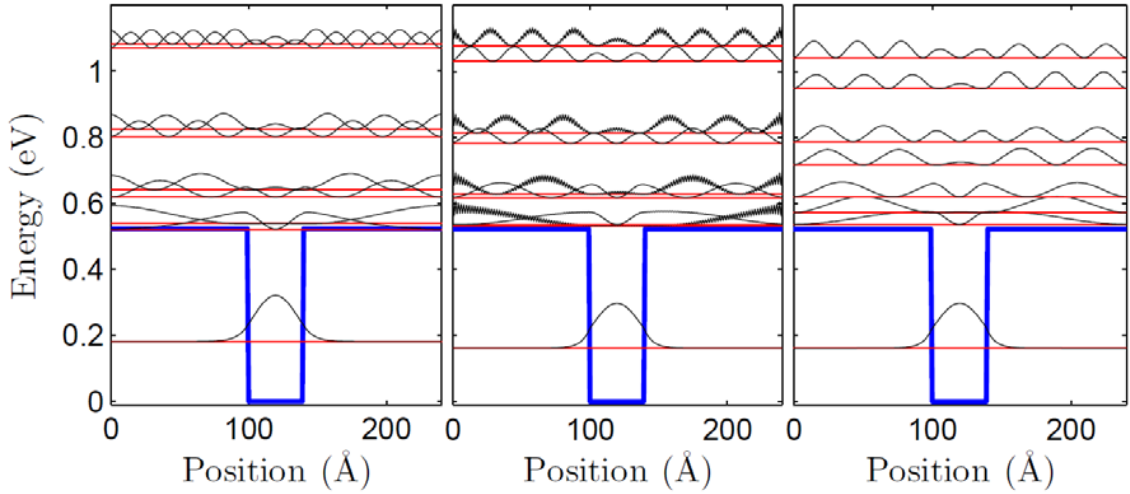

FIG. 4. The eigen solutions in a 40 Å $Ga_{0.47}In_{0.53}As$ / $Al_{0.48}In_{0.52}As$ single quantum well are compared for three different approaches described in section II based on the one band model. Left panel: the Fourier grid Hamiltonian method (FGHM), middle panel: the modified Fourier grid Hamiltonian method (MFGHM), right panel: the delta function method (DFM). In the FGHM and the DFM, spurious solutions do not appear. However, the MFGHM produces them in the continuum as fast oscillating envelope functions. The spurious solutions are responsible for the difference of continuum states between the MFGHM and the DFM.

Since the FGHM does not produce unphysical solutions, we were able to figure out their origin and remove them by comparing with the MFGHM. The only difference between the two methods lies in the manner how the quasi-particle wave number integrals are dealt with. The FGHM treats them by the discretization with equal lengths as in Eq. (38), while the MFGHM exactly evaluates them analytically as in Eq. (39).

$$\frac{2\pi}{N}(\Delta k)\sum_{\beta=1}^{m}\beta\cos\left[\frac{2\pi\beta(p-s)}{N}\right] \quad (38)$$

$$\frac{k_m \cos\left(\frac{2\pi}{N}m(p-s)\right)}{(p-s)} - \frac{\sin\left(\frac{2\pi}{N}m(p-s)\right)}{(p-s)^2 \Delta z} \quad (39)$$

where $\Delta z$ is taken to be 1 Å for simplicity. In (39), the first cosine term is dominant due to the factor $k_m$.

For a 40 Å GaInAs/AlInAs well, expressions (38) and (39) are plotted in FIG. 5 as a function of the index s (see Eq. (29, 30)) for a given p = 120, which corresponds to the center of the quantum well. Each of (38) and (39) is an even or an odd function, which shows a peak or a zero point at |p - s| = 0 respectively, and the parity becomes opposite at that point. Also, as the absolute value |p - s| increases, the amplitudes of (38) are much more quickly suppressed than those of (39). It is important to notice that their different behaviors are not related to the cut-off value of wave vector k since the latter is determined by the reciprocal length of a heterostructure in both cases.

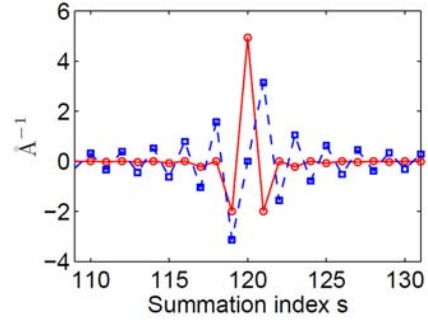

FIG. 5. The wave number integrals, expression (38) (circle solid line) and (39) (square, dashed line) that appear in the Fourier grid Hamiltonian method (FGHM) (see (8)) and the modified Fourier grid Hamiltonian method (MFGHM) (see (12)) respectively are plotted as a function of the summation coordinate index s for a given real space basis p = 120, which corresponds to the center of the quantum well shown in FIG. 4 with the grid length of 1 Å. The parity between the two cases becomes opposite at s = 121 and higher.

To remove the spurious solutions in the MFGHM, we try to achieve the characteristics of the wave number integral in the FGHM by introducing a certain shift factor α, i.e., with replacing s to s-α in (39). Any small shift factor can lead to a non-zero wave number integral at |p - s| = 0 (or |q - s| = 0). Also, the shift of Δz/2 makes the wave number integral decay much more quickly as |p - s| (or |q - s|) increases. Such behavior originates from the destructive interference of the wave number integral due to the shift. Figure 6 shows expression (39) before and after applying the shift factor.

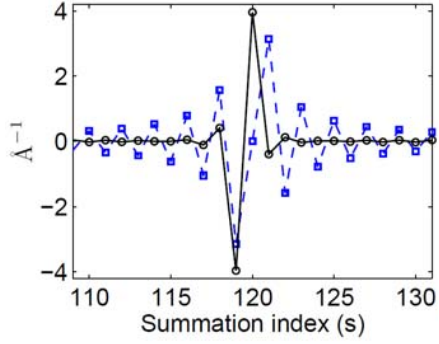

FIG. 6. The wave number integral, expression (39), before (square, dashed line) and after (circle, solid line) adding a shift factor $+\Delta z/2$ to s in (39) calculated by the MFGHM for a given p = 120. Such a shift factor removes the fast oscillating spurious solutions in the continuum in the MFGHM as shown in FIG. 7, which can be compared to FIG. 4 (middle panel) before the removal of fast oscillating spurious solutions.

In addition, to keep the symmetry of envelope functions even after introducing the shift factor, the average of two Hamiltonians obtained by the positive and negative shifts is used. The resulting eigen solutions are now free from the spurious solutions as shown in FIG. 7 with the shift factor $\alpha = \pm\Delta z/2$, and they resemble those obtained by the DFM in FIG. 4 (right) rather than those by the FGHM (FIG. 4 (left)). The spurious-solution-free eigenvalues after the shift factor is introduced in the MFGHM are compared with those obtained by the DFM in TABLE III for the same quantum well structure as shown in FIG. 4 with the same grid length of 1 Å. Subband levels in the DFM are more accurate now. Since the larger shifts diminish the accuracy of eigen solutions, the minimal shift to just remove the spurious solutions will be optimal. The removal of fast oscillating envelope functions comes at the price of approximating the exact wave number integral.

Note that the removal of spurious solutions is not attributed to the implicit change of interface boundary conditions due to the shift factor. We will extend the discussion on the spurious solutions for the multiband case in the next section.

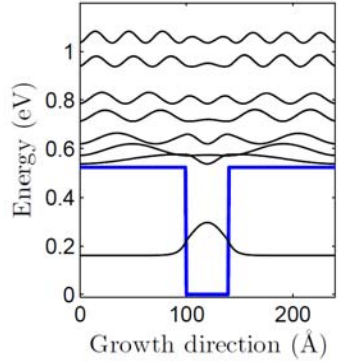

FIG. 7. The eigenstates (envelope functions) obtained by the MFGHN after introducing the shift factor of $\pm\Delta z/2$ in (39). The fast oscillating envelope functions in the continuum present in FIG. 4 (middle panel) have now disappeared. The average of Hamiltonians for each positive and negative shift is used to preserve the symmetry of envelope functions based on the one-band model. The subband levels now resemble those obtained by the DFM. (see FIG. 4 (right panel)) Their numerical values are compared in TABLE III for the same grid length of 1 Å. A small shift factor weakly affects confined energy levels.

TABLE III. Energies of confined states in a single 40 Å $Ga_{0.47}In_{0.53}As$ quantum well surrounded by a 100 Å $Al_{0.48}In_{0.52}As$ barrier, after spurious solution have been removed by introducing the shift factor $\pm\Delta z/2$ in wave number integrals (see the text) in the MFGHM are compared to eigen solutions obtained in the DFM.

|       | Spurious-solution-free MFGHM (meV) | DFM (meV) |
|-------|------------------------------------|-----------|
| CB 1  | 162.361                            | 161.148   |
| CB 2  | 536.142                            | 536.526   |
| CB 3  | 570.545                            | 572.130   |
| CB 4  | 617.174                            | 620.103   |
| CB 5  | 711.380                            | 717.262   |
| CB 6  | 780.046                            | 787.767   |

### B. Heterostructure eigen solutions in multiband k • p models

The Hamiltonians constructed in section III for heterostructure problems with one dimensional confinement based on multiband k • p models (see the Appendix for the bulk Hamiltonian of a 8-band k • p model) can be easily solved for eigen solutions by a standard eigenvalue solver. The numerical results for confined subband energy levels in a single GaAs/AlGaAs quantum well by the 6-valence band and the 8-band k • p models are shown in TABLE IV for the in-plane wave vector $k_\parallel = 0$. In the 6-valence band case, the subband positions obtained by the FGHM significantly deviate from those obtained by the other two methods, i.e., the MFGHM and the DFM. The discrepancy becomes larger with increasing $k_\parallel$ as shown in FIG. 8 and 9. On the other hand, the MFGHM and the DFM give nearly the same subband positions. FIG. 8 shows such an agreement in the in-plane subband dispersion. Note that the DFM and the FGHM do not produce any kind of spurious solutions within the 6-valence band model. On the other hand, in the MFGHM they occur in the continuum again and can be removed just like in its one band model case. The general trend of eigen solutions obtained by the three methods in the 6-valence k • p band model is consistent with that in the one band model.

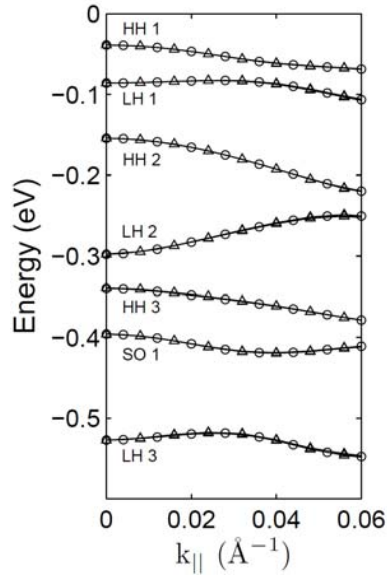

FIG. 8. The in-plane dispersion of subbands in the valence band of a 42 Å GaAs/AlAs single quantum well obtained by the MFGHM (circle) and the DFM (triangle) based on the 6-valence band k • p model. Energies are measured from the top of the valence band. Results can be compared to FIG. 5 in Ref. 8.

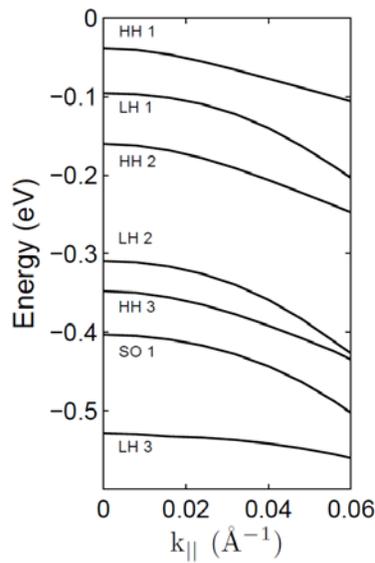

FIG. 9. The in-plane dispersion of subbands in the valence band of a 42 Å GaAs/AlAs single quantum well heterostructure obtained by the FGHM based on the 6-valence band k • p model. Energies are measured from the top of the valence band. This plot shows a severe discrepancy with FIG. 8 as $k_\parallel$ becomes large.

TABLE IV. The subband levels obtained by the three heterostructure Hamiltonian methods are compared within the 6-valence band model and the 8-band k • p model for a single 50 Å GaAs/Ga$_{0.3}$Al$_{0.7}$As quantum well at $k_\parallel = 0$. The differences in the subband levels between the MFGHM and the DFM are less than 1 meV in both models. As with the one band model, the eigen solutions obtained by the FGHM are significantly different from those obtained by the other two methods. Here a grid length of 1 Å has been used.

|  | FGHM (meV) | MFGHM (meV) | DFM (meV) |
|---|---|---|---|
| | 8-band k • p model | | |
| CB1 | 108.645 | 107.250 | 107.156 |
| CB2 | 376.827 | 370.372 | 369.781 |
| HH1 | -27.697 | -27.889 | -27.861 |
| LH1 | -72.154 | -70.371 | -70.329 |
| HH2 | -113.081 | -110.411 | -110.283 |
| LH2 | -232.845 | -227.900 | -227.663 |
| HH3 | -245.384 | -241.840 | -241.450 |
| SO1 | -360.557 | -360.655 | -361.282 |
| | 6-valence band k • p model | | |
| HH1 | -27.697 | -27.889 | -27.861 |
| LH1 | -70.431 | -66.448 | -66.415 |
| HH2 | -113.081 | -110.412 | -110.283 |
| LH2 | -237.240 | -231.095 | -230.985 |
| HH3 | -245.384 | -241.840 | -241.450 |
| SO1 | -360.728 | -360.892 | -361.526 |

Upon including the conduction band to the 6-valence band, i.e., in the 8-band model, subband levels at $k_{\parallel} = 0$, calculated by the FGHM, the MFGHM, and the DFM, are compared in TABLE IV. The general pattern in terms of accuracy of eigen solutions is the same as in both the one-band and the 6-valence band cases. However, in the transition from the 6-band to the 8-band model, unphysical solutions of another kind appear in the band gap in the DFM as shown in FIG. 10 (right). Also, the fast oscillating envelope functions can occur in very high continuum states even within the DFM although they are not shown in the chosen quantum well heterostructure in FIG. 10~12. The tendency of spurious solutions in a heterostructure calculated by the MFGHM under the 8-band k • p model is more or less the same as in both the one- band and the 6-valence band models, showing the fast oscillating continuum states and no spurious solutions in confined states. The eigen solutions obtained by the FGHM are still completely free from any kind of spurious solutions even if they are not so accurate. The above features of the three different approaches in the real space heterostructure Hamiltonian methods are summarized in TABLE V.

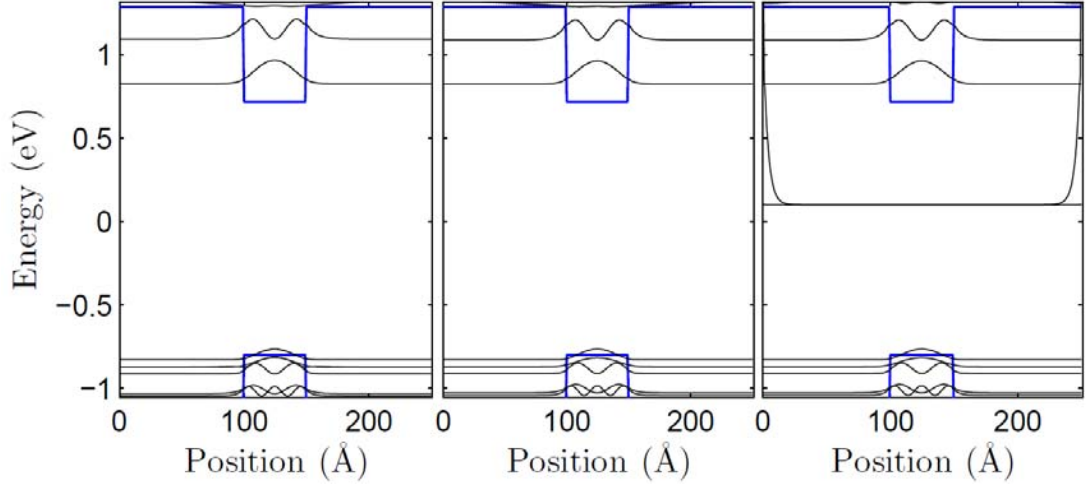

FIG. 10. Eigen solutions in a single quantum well GaAs/Al$_{0.7}$Ga$_{0.3}$As heterostructure of well width 50 Å, obtained by the FGHM (left), the MFGHM (middle), and the DFM (right) based on the 8-band k • p model are shown near the band gap region. The first two heterostructure Hamiltonian methods do not produce any spurious solutions in the band gap, but the DFM does.

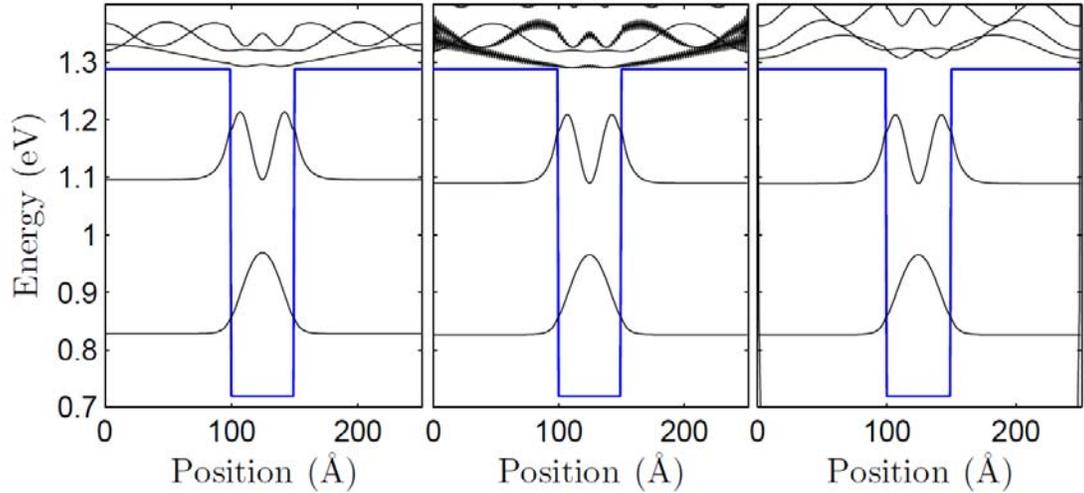

FIG. 11. Eigen solutions in the conduction band of a single quantum well GaAs/Al$_{0.7}$Ga$_{0.3}$As heterostructure of well width 50 Å, obtained by the FGHM (left), the MFGHM (middle), and the DFM (right) at the Γ point based on the 8-band k • p model. Only the MFGHM produces fast

oscillating envelope functions in the continuum. However, such spurious solutions do not occur in the confined states. The amplitudes of envelope functions have been enhanced for better visualization.

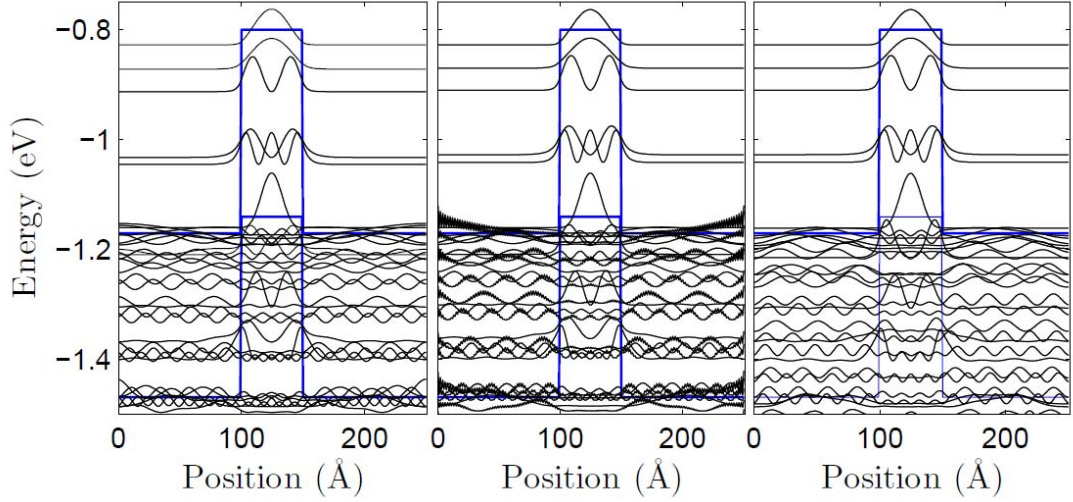

FIG. 12. Eigen solutions in the valence band of a single quantum well GaAs/Al$_{0.7}$Ga$_{0.3}$As heterostructure of well width 50 Å, obtained by the FGHM (left), the MFGHM (middle), and the DFM (right) at the $\Gamma$ point based on the 8-band k • p model. Only the MFGHM produces the fast oscillating envelope functions in the continuum. However, such spurious solutions do not occur for the confined states. The amplitudes of envelope functions have been enhanced for better visualization.

When comparing the eigen solutions obtained by the 6- and 8-band models in TABLE IV, it is important to recognize that the occurrence of spurious solutions in the middle of the band gap does not affect the true confined eigen states.

In the case of the 8-band model, we apply the same strategy that was used for eliminating fast oscillating continuum states in the one-band and the 6-valence band model by introducing the shift factor $\pm\Delta z/2$ in the wave number integral in the MFGHM followed by averaging heterostructure Hamiltonians at positive and negative shifts. As a result, most of the fast oscillating envelope functions in the valence band continuum are removed, but those in the conduction band continuum still reside nearly without change. However, note that there are no unphysical solutions in the band gap as in the one- and 6-valence band cases.

TABLE V. The characteristics of the three approaches in the heterostructure Hamiltonian method are compared regarding the generation of spurious solutions and the accuracy of eigen solutions in the one-band, the 6-valence band, and the 8-band k • p models.

|  | One-band model | | 6-valence band k • p model | | 8-band k • p model | |
|---|---|---|---|---|---|---|
|  | Generation of spurious solutions | Accurate confined eigen-solutions | Generation of spurious solutions | Accurate confined eigen-solutions | Generation of spurious solutions | Accurate confined eigen-solutions |
| FGHM | No | No | No | No | No | No |
| MFGHM | Only in the continuum | Yes | Only in the continuum in CB & VB | Yes | Only in the continuum in CB & VB | Yes |
| DFM | No | Yes | No | Yes | In the high continuum of CB or/and in BG | Yes |

To summarize the behavior of the spurious solutions, the unphysical eigen solutions in the middle of the band gap are originated from the contribution of large wave number values in the wave number integrals in sections II. B and II. C. In the MFGHM, upon evaluating the integrals, k values have been cut off by the Fourier reciprocal relation. On the other hand, in the DFM, those integrals have been analytically integrated over the infinite range without any truncation of k, and the final formalism solely depends on the band parameters in the real space. For the other type of spurious solutions, i.e., the fast oscillating envelope functions, their removal is related to sacrificing the accuracy of the eigen solutions as can be seen in the comparison of the FGHM and the MFGHM.

The in-plane subband dispersion in the valence band, obtained by the MFGHM, are compared for the 6-and 8-band cases in FIG. 13, i.e., with and without an explicit inclusion of the conduction band, which is done by modifying the Luttinger parameters. It shows that the change of the valence band interaction with remote bands in bulk materials influence the valence band in-plane dispersion in heterostructures. Figure 13 can be compared to FIG. 5 in Ref.[8].

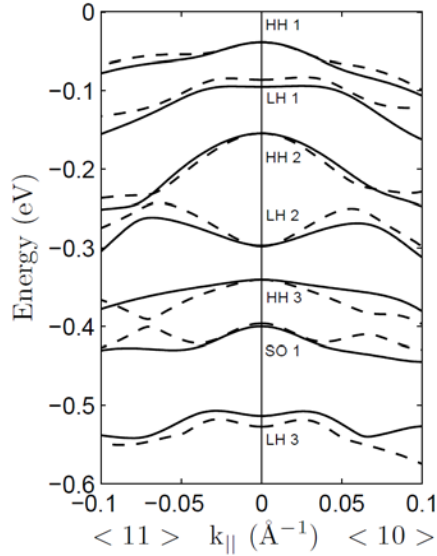

FIG. 13. The in-plane dispersion of subbands in the valence band of a 42 Å GaAs/AlAs single quantum well heterostructure obtained by the MFGHM based on the 6-valence band (dashed) and the 8-band (solid) k • p model. At large $k_\parallel$, the disagreement becomes obvious. This indicates that the modification of the valence band interaction parameters (Luttinger parameters) due to the explicit inclusion of the conduction band has non-negligible influences on subband positions. The crossing of subband dispersions is sharper in the 6-band model.

We also investigated the effect of square wave-like abrupt hetero-interfaces of bulk band parameters on spurious solutions by the Fourier series expansion[41] as in expression (40). The abruptness of the band parameters at interfaces is effectively controlled by the upper limit, $n_{max}$, of the summation in (40). Figure 14 shows that such replacement only improves the smoothness of envelope functions at the interface boundary, and it rarely affects both types of spurious solutions.

$$B(z) \approx \frac{a_0}{2} + \sum_{n=1}^{n_{max}} a_n \cos(nk_L z) + \sum_{n=1}^{n_{max}} a_n \sin(nk_L z) \tag{40}$$

where $k_L = 2\pi/L$, L is a heterostructure length, and

$$a_0 = \frac{2}{L} \sum_{s=1}^{N_{ly}} B_s (L_{s+1} - L_s)$$

$$a_n = \frac{1}{n\pi} \sum_{s=1}^{N_{ly}} B_s [\sin(nk_L L_{s+1}) - \sin(nk_L L_s)]$$

$$b_n = -\frac{1}{n\pi} \sum_{s=1}^{N_{ly}} B_s [\cos(nk_L L_{s+1}) - \cos(nk_L L_s)]$$

where $N_{ly}$ is the number of layers, $L_1 = 0$, and $L_{N_{ly}+1} = L$.

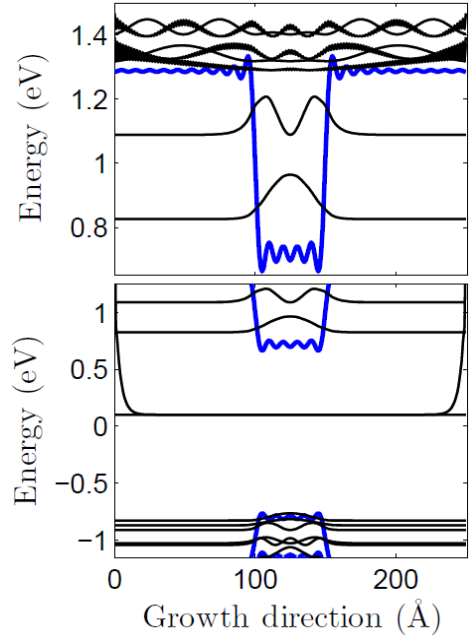

FIG. 14. The same quantum well as in FIG. 13 but with square wave-like abrupt interfaces of band parameters replaced by the smooth ones using the Fourier expansion of the coordinate dependent band parameters, Eq. (37), calculated by the MFGHM (top) and the DFM (bottom). Smoothness of interfaces does not affect the spurious solutions in both cases, only the envelope functions at interfaces become smoother. There are 9 interface grid points with $n_{max} = 25$ in Eq. (37).

Figure 15 shows the eigen solutions in a single GaAs layer calculated by the DFM, in which band parameters in Ref.[39] are used, and a spurious solution is observed. As long as such spurious solutions appear in a single GaAs layer, it is generally impossible to completely remove them in GaAs-based heterostructures even if some of heterostructures can be free from spurious solutions due to certain destructive interferences. The way to modify the Luttinger parameters to prevent spurious solutions in the band gap has been reported based on the finite difference method.[14]

However, this prescription fails in the general case. The universal method which can remove the existing spurious solutions within the EFA is currently absent.

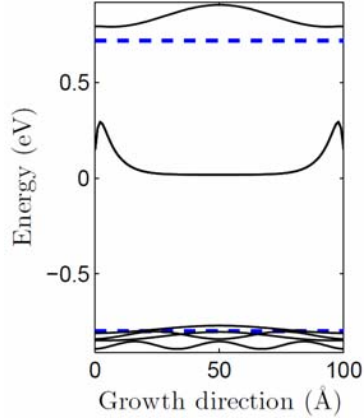

FIG. 15. Eigen solutions obtained by the DFM with the 8-band k • p model in a 100 Å GaAs single layer. The unphysical eigen solutions inside the band gap (dashed) can be observed. The band parameters in Ref. 39 are used; the band parameters recommended in Ref. 14 cannot remove the solution in the middle of the band gap.

## V. SUMMARY AND CONCLUSION

We presented the heterostructure Hamiltonian method in the real space with three different approaches for calculating eigen states of layered semiconductor heterostructures. They have been tested in the one-band, the 6-valence band, and the 8-band k • p models and they can be applied to general n-band k • p models within the envelope function approximation. Inherent advantages of the method include the treatment of the boundary conditions that are automatically satisfied, extreme simplicity, transparency, the high accuracy of true eigen solutions, and the unified explanation and removal of spurious solutions.

## ACKNOWLEDGEMENTS

This work was supported in part by NSF grants ECS-0547019, EEC-0540832 and ECCS-0925446, and by the NHARP project 003658-0010-2009.

## APPENDIX: THE 8×8 k • p HAMILTONIAN [23]

In the non-relativistic limit, the Hamiltonian has the form as shown in (A. 1), in which the spin-orbit interaction term is added to the Schrödinger equation and with bare electron mass m. The effect of the spin-orbit interaction becomes more pronounced for heavier semiconductors.

$$H_0 = \frac{\mathbf{p}^2}{2m} + V(x) + \frac{\hbar}{4m^2c^2}(\boldsymbol{\sigma} \times \nabla V) \bullet \mathbf{p} \tag{A. 1}$$

After substituting the Bloch wavefunction $\psi_{n\mathbf{k}} = \frac{1}{\sqrt{V}} e^{i\mathbf{k}\bullet\mathbf{r}} u_{n\mathbf{k}}$ to (A. 1), only the periodic part of the Bloch function is left, cancelling the plane wave part in $H_0 \psi_{n\mathbf{k}} = E_{n\mathbf{k}} \psi_{n\mathbf{k}}$. Consequently, it gives the k-dependent Hamiltonian as in (A. 2) and $H_\mathbf{k} u_{n\mathbf{k}} = E_{n\mathbf{k}} u_{n\mathbf{k}}$.

$$H_\mathbf{k} = \frac{\mathbf{p}^2}{2m} + V(x) + \frac{\hbar^2 k^2}{2m} + \frac{\hbar}{m}\mathbf{k}\bullet\mathbf{p} + \frac{\hbar}{4m^2c^2}(\boldsymbol{\sigma}\times\nabla V)\bullet\mathbf{p} + \frac{\hbar^2}{4m^2c^2}(\boldsymbol{\sigma}\times\nabla V)\bullet\mathbf{k} \tag{A. 2}$$

$$= \frac{\mathbf{p}^2}{2m} + V(x) + \frac{\hbar}{4m^2c^2}(\boldsymbol{\sigma}\times\nabla V)\bullet\mathbf{p} + \frac{\hbar^2 k^2}{2m} + \frac{\hbar}{m}\mathbf{k}\bullet\left[\mathbf{p} + \frac{\hbar^2}{4m^2c^2}(\boldsymbol{\sigma}\times\nabla V)\right] \tag{A. 3}$$

Using $\boldsymbol{\pi} \equiv \mathbf{p} + \frac{\hbar^2}{4m^2c^2}(\boldsymbol{\sigma}\times\nabla V)$ and (A. 1), expression (A. 3) is simplified as

$$H_\mathbf{k} = H_0 + \frac{\hbar^2 k^2}{2m} + \frac{\hbar}{m}\mathbf{k}\bullet\boldsymbol{\pi} \tag{A. 4}$$

Using the perturbation theory, $E_{n\mathbf{k}}$ and the periodic part of the Bloch function $u_{n\mathbf{k}}$ can be expanded up to the second order and the first order respectively with respect to small $\mathbf{k}$. Using the zeroth order of the eight periodic parts (A. 5) of the Bloch function that are determined by the symmetry of orbitals, corresponding to $\Gamma_{6C}$(CB), $\Gamma_{8V}$(HH, LH) and $\Gamma_{7V}$(SO) (the double point group notation) in the $\Gamma$ symmetry point, in which the Kramer's degeneracy is taken into account, the resulting Hermitian Hamiltonian becomes the 8×8 $\mathbf{k} \cdot \mathbf{p}$ Hamiltonian as shown in (A. 6). The higher bands that are cut off in this band model are effectively included in the second order terms through the Luttinger parameters $\gamma_{1,2,3}$ and the Kane parameter F.

$$|1/2,1/2\rangle_c = |S\uparrow\rangle, \quad |1/2,-1/2\rangle_c = |S\downarrow\rangle,$$

$$|3/2,3/2\rangle = \frac{-i}{\sqrt{2}}|(X+iY)\uparrow\rangle, \quad |3/2,-3/2\rangle = \frac{i}{\sqrt{2}}|(X-iY)\downarrow\rangle,$$

$$|3/2,1/2\rangle = \frac{i}{\sqrt{6}}\left[-|(X+iY)\downarrow\rangle + 2|Z\uparrow\rangle\right], \quad |3/2,-1/2\rangle = i\left[\frac{1}{\sqrt{6}}|(X-iY)\uparrow\rangle + \sqrt{\frac{2}{3}}|Z\downarrow\rangle\right],$$

$$|1/2,1/2\rangle = i\left[\frac{1}{\sqrt{3}}|(X+iY)\downarrow\rangle + \sqrt{\frac{1}{3}}|Z\uparrow\rangle\right], \quad |1/2,-1/2\rangle = \frac{i}{\sqrt{3}}\left[|(X-iY)\uparrow\rangle - |Z\downarrow\rangle\right] \quad \text{(A. 5)}$$

$$H_{8\times8} = \begin{pmatrix}
 & (1/2,1/2) & (1/2,-1/2) & (3/2,3/2) & (3/2,1/2) & (3/2,-1/2) & (3/2,-3/2) & (1/2,1/2) & (1/2,-1/2) \\
(1/2,1/2) & E_g + \frac{\hbar^2 k^2}{2m_0}(1+2F) + \phi_E & 0 & -Pk_+ & \sqrt{\frac{2}{3}}Pk_z & \frac{1}{\sqrt{3}}Pk_- & 0 & \frac{1}{\sqrt{3}}Pk_z & \sqrt{\frac{2}{3}}Pk_- \\
(1/2,-1/2) & 0 & E_g + \frac{\hbar^2 k^2}{2m_0}(1+2F) + \phi_E & 0 & \frac{-1}{\sqrt{3}}Pk_+ & \sqrt{\frac{2}{3}}Pk_z & Pk_- & \sqrt{\frac{2}{3}}Pk_+ & \frac{-1}{\sqrt{3}}Pk_z \\
(3/2,3/2) & -Pk_- & 0 & F'+\phi_E & H' & I' & 0 & \frac{H'}{\sqrt{2}} & \sqrt{2}I' \\
(3/2,1/2) & \sqrt{\frac{2}{3}}Pk_z & \frac{-1}{\sqrt{3}}Pk_- & H^{*\prime} & G'+\phi_E & 0 & I' & \frac{1}{\sqrt{2}}(G'-F') & -\sqrt{\frac{3}{2}}H' \\
(3/2,-1/2) & \frac{1}{\sqrt{3}}Pk_+ & \sqrt{\frac{2}{3}}Pk_z & I^{*\prime} & 0 & G'+\phi_E & -H' & -\sqrt{\frac{3}{2}}H^{*\prime} & \frac{-1}{\sqrt{2}}(G'-F') \\
(3/2,-3/2) & 0 & Pk_+ & 0 & I^{*\prime} & -H^{*\prime} & F'+\phi_E & -\sqrt{2}I^{*\prime} & \frac{1}{\sqrt{2}}H^{*\prime} \\
(1/2,1/2) & \frac{1}{\sqrt{3}}Pk_z & \sqrt{\frac{2}{3}}Pk_- & \frac{H^{*\prime}}{\sqrt{2}} & \frac{1}{\sqrt{2}}(G^{*\prime}-F^{*\prime}) & -\sqrt{\frac{3}{2}}H' & -\sqrt{2}I' & -\Delta+\frac{F'+G'}{2}+\phi_E & 0 \\
(1/2,-1/2) & \sqrt{\frac{2}{3}}Pk_+ & \frac{-1}{\sqrt{3}}Pk_z & \sqrt{2}I^{*\prime} & -\sqrt{\frac{3}{2}}H^{*\prime} & \frac{-1}{\sqrt{2}}(G^{*\prime}-F^{*\prime}) & \frac{1}{\sqrt{2}}H' & 0 & -\Delta+\frac{F'+G'}{2}+\phi_E
\end{pmatrix}$$

(A. 6)

where $E_g$ is the band gap and $\Delta$ is the spin orbit split-off energy, $\phi_E$ is the potential energy due to an external electric field, $k_\pm = (1/\sqrt{2})(k_x \pm i k_y)$, and other parameters are defined as follows;

$$F' = -\frac{\hbar^2}{2m_0}\left[(\gamma_1 + \gamma_2)k_\parallel^2 + (\gamma_1 - 2\gamma_2)k_z^2\right], \quad G' = -\frac{\hbar^2}{2m_0}\left[(\gamma_1 - \gamma_2)k_\parallel^2 + (\gamma_1 + 2\gamma_2)k_z^2\right],$$

$$H' = \frac{\hbar^2}{m_0}\sqrt{3}\gamma_3 k_z(k_x - ik_y) = \frac{\hbar^2}{2m_0}2\sqrt{6}\gamma_3 k_z k_-, \quad I' = \frac{\hbar^2}{2m_0}\left[\sqrt{3}\gamma_2(k_x^2 - k_y^2) - i2\sqrt{3}\gamma_3 k_x k_y\right]$$

$$P = \frac{i\hbar}{m_0}\langle S|p_x|X\rangle \equiv \frac{i\hbar}{m_0}p_{SX}^x$$

$$\gamma_1 = \gamma_1^L - \frac{2m_0}{3\hbar^2}\frac{P^2}{E_g}, \quad \gamma_{2,3} = \gamma_{2,3}^L - \frac{m_0}{3\hbar^2}\frac{P^2}{E_g},$$

$$\gamma_1^L = -1 - \frac{2}{3m_0}\sum_r \frac{p_{Xr}^x p_{rX}^x}{\varepsilon_0 - \varepsilon_r} - \frac{4}{3m_0}\sum_r \frac{p_{Xr}^y p_{rX}^y}{\varepsilon_0 - \varepsilon_r},$$

$$\gamma_2^L = -\frac{1}{3m_0}\left(\sum_r \frac{p_{Xr}^x p_{rX}^x}{\varepsilon_0 - \varepsilon_r} - \sum_r \frac{p_{Xr}^y p_{rX}^y}{\varepsilon_0 - \varepsilon_r}\right),$$

$$\gamma_3^L = -\frac{1}{3m_0}\left(\sum_r \frac{p_{Xr}^x p_{rX}^x + p_{Yr}^y p_{rY}^y}{\varepsilon_0 - \varepsilon_r}\right)$$

where $m_0$ is the bare electron mass, $\hbar$ is the Plank constant, $p_\alpha$ is the $\alpha$-component of the momentum, $k_\parallel^2 = k_x^2 + k_y^2$, $\varepsilon_0$ is the band edge energy for corresponding band, originated from $H_0$ in (A. 4), S is the s-like orbital and the linear combination of X, Y, and Z is the p-like orbital, $\varepsilon_{rc}$ and $\varepsilon_r$ represent all CB edge energies higher than the first CB and all remote band edge energies respectively.

In the Luttinger parameters $\gamma_{1,2,3}^L$, the free carrier energy term $\hbar^2 k^2/2m_0$ in (A. 4) is already included. Since the CB is explicitly included in the band model, its contributions are subtracted from the Luttinger parameters $\gamma_{1,2,3}^L$, giving the modified Luttinger parameters $\gamma_{1,2,3}$.

# REFERENCES

*belyanin@tamu.edu